\documentclass[conference]{IEEEtran}
\IEEEoverridecommandlockouts
\usepackage{cite}
\usepackage{amsmath,amssymb,amsfonts}
\usepackage{graphicx}
\usepackage{textcomp}
\usepackage{xcolor}
\usepackage{algorithm}
\usepackage{algpseudocode}

\usepackage[utf8]{inputenc}
\usepackage{multirow}
\usepackage{graphicx}

\def\BibTeX{{\rm B\kern-.05em{\sc i\kern-.025em b}\kern-.08em
    T\kern-.1667em\lower.7ex\hbox{E}\kern-.125emX}}
\begin{document}

\title{Optimal partitioning in distributed state estimation considering a modified convergence criterion\\
}

\author{\IEEEauthorblockN{Sajjad Asefi}
\IEEEauthorblockA{\textit{Center for Energy Science \& Technology} \\
\textit{Skolkovo Institute of Science \& Technology}\\
Moscow, Russia \\
sajjad.asefi@skoltech.ru}
\and
\IEEEauthorblockN{Elena Gryazina}
\IEEEauthorblockA{\textit{Center for Energy Science \& Technology} \\
\textit{Skolkovo Institute of Science \& Technology}\\
Moscow, Russia \\
e.gryazina@skoltech.ru}
\and
\IEEEauthorblockN{Helder Leite}
\IEEEauthorblockA{\textit{Electrical Engineering department} \\
\textit{University of Porto}\\
Porto, Portugal \\
hleite@fe.up.pt}
}

\maketitle

\begin{abstract}
Distributed state estimation (DSE) is considered as a more robust and reliable alternative for centralized state estimation (CSE) in power system. Especially, taking into account the future power grid, so called smart grid in which bi-directional transfer of energy and information happens, and renewable energy sources with huge indeterminacy are applied more than before. Combining the mentioned features and complexity of the power network, there is a high probability that CSE face problems such as communication bottleneck or security/reliability issues. So, DSE has the potential to be considered as a solution to solve the mentioned issues. In this paper, first, a modified convergence criterion is proposed and has been tested for different approaches of DSE problem, considering the most important factors such as iteration number, convergence rate, and data needed to be transferred to/from each area. Then, an optimal partitioning technique has been implemented for clustering the system into different areas. Besides the detailed analysis and comparison of recent DSE methods, the proposed partitioning method's effectiveness has been shown in this paper.
\end{abstract}


\section{Introduction} \label{sec: intro}

State estimation (SE) is an indispensable part of modern power systems. The significant advantages of SE are the estimation of network parameters (i.e., voltage magnitude and phase angle) based on redundancy in measurements, bad data detection, estimating delayed or missing data \cite{wood2013power, zhao2020roles}. Improved SE can enhance monitoring and control of the power system in the state of contingency. State estimator's primary responsibility is to provide reliable and complete information, which has great importance for operations and control systems that ensure the power grid's security \cite{gomez2018electric}. In other words, the state estimator's main role is to identify the system state by minimizing a specific criterion based on up-to-date system measurements \cite{ahmad2018distribution}. Also, the expanding presence of distributed energy resources (DERs) in the power grid demands SE to be more precise and fast due to DERs' intermittent nature.

Power system SE problem has been proposed to be applied from 1970s, when first Schweppe et al. realized a model to solve this issue \cite{schweppe1970power1, schweppe1970power2, schweppe1970power3}. Traditionally SE is performed in a centralized manner, in which an individual unit collects all input data (‘data’ here relates to measurement unit values and system parameters such as line impedance) of the system, and an optimization method is used to solve the SE problem. Later on, not only for SE but also for other optimization problems, distributed solutions attracted researchers interest. Despite of the fact that there have been many researches about Distributed state estimation (DSE) but not all the technical details have been examined in the literature.

In \cite{conejo2007optimization}, a general multi-area SE method is illustrated, providing an accurate estimation of the system states. It is to be noted that the areas interchange only a small amount of border data, needless to be processed or manipulated. Authors in \cite{kekatos2013decentralized} and \cite{kekatos2013distributed} have provided a novel algorithm for DSE using the alternating direction method of multipliers (ADMM). In \cite{minot2015fully} and \cite{minot2016distributed}, a matrix splitting based DSE method is presented. In \cite{minot2015fully}, authors introduced the technique for DC SE, but later in \cite{minot2016distributed} they showed a distributed Gauss-Newton method for AC SE.

Evidently, any optimization procedure needs certain convergence criterion. Usually this is done considering an accuracy limit of objective function, step tolerance and optimality tolerance. Decentralization decreases the possibility of communication bottleneck, which is more probable to happen for a huge system centralized state estimation (CSE) \cite{kekatos2017psse}. Additionally, in some cases due to data privacy and cybersecurity reasons, the SE must be distributed, e.g., when areas (‘area’ here refers to a partition of the power system) are operated by different RTOs. In contrast to CSE, novel modified convergence criterion should be proposed and applied so as to reduce the amount information sent through the communication channel. For application of distributed methods we need to partition the power system into different areas to increase the ability of the system to withstand the risks \cite{du2019admm}. As an example, in \cite{irving1990optimal} an algorithm based on stimulated annealing principle has been presented, in order to divide a power system into smaller sub-systems that will help parallel computers to analyze the system. 

In this paper, we examined application of proposed modified convergence criterion on recent and well-known DSE methods by IEEE standard test systems, and the performance of these methods from different viewpoints, such as data needed to be transferred, computation time, error of  obtained solution compared to centralized one and convergence rate, are compared and the best method is selected. This comparison serves for the main goal of the paper which is application of an optimal partitioning method which decreases communication burden and needed data to be transferred to/from each area and consequently, decreases iteration number for getting a reasonable solution. To best of the authors’ knowledge, the detailed analysis of DSE methods and the effect of convergence criterion and optimal partitioning on DSE features has not been addressed in the literature.

The rest of the article is composed of the following sections. Section \ref{sec: prob-form} discusses about the main problem formulation and objective function. Additionally, studied DSE methods are briefly presented in this section. Section \ref{sec: results} presents the results obtained by simulation, plus a brief discussion about them. And eventually, in Section \ref{sec: conclusion} the paper is concluded.

\section{Problem statement and formulation} \label{sec: prob-form}
 In this section, the mathematical equations governing the SE problem is presented. After, providing the general problem formulation of SE, a brief overview about different DSE known approaches are provided. And at the end, convergence criterion and the system partitioning formulation is provided.
 
\subsection{Distributed state estimation}
The type of SE we have considered in this work is static SE. Static SE is used to keep track of the network during quasi steady-state responding to slowly varying network load and generation. It is possible to express the relation between system states, i.e., voltages' magnitude and phase angle, and all measurements, such as power flows, power injections, voltage magnitudes, etc., as a minimization problem. Based on the concept of maximum likelihood and assuming the independent measurement value errors which follow the normal probability distribution function \cite{gomez2018electric}, SE problem can be written in the following format: 
\begin{equation}
 min \text{ } [z - f(x)]^T \times R^{-1} \times [z-f(x)]
\label{eq: main_objective_function}
\end{equation}
where $R$ is the diagonal covariance matrix related to the measurements, $z$ is the measurements matrix and $f(x)$ is a function that shows the relation between measurements and state variables. There are two different viewpoint to SE problem in general. One is DC SE, in which the $f(x)$ is a linear function (that means measurements have linear relation with state variables) and state variables are only voltage phase angles ($\theta$s). The other one is AC SE, in which $f(x)$ is nonlinear and need a linearization step to solve it. 

DC SE has been a matter of concern in this paper. If we consider a linear relation between measurement units and state variables ($f(x)=F\times x$, where $F$ indicates the measurement matrix and $x$ represents all $\theta$s of the power system), the final solution to \eqref{eq: main_objective_function}, $x^*$, can be obtained as follows, meeting the first order optimality condition:

\begin{equation}
\begin{matrix}
    ({F^T}R^{-1}F)x={F^T}R^{-1}z\\[0.5 cm]
    x^* = ({F^T}R^{-1}F)^{-1}\times {F^T}R^{-1}z
\end{matrix}
\label{eq: centralized_eq}
\end{equation}

It is to be noted that to have a solution for \eqref{eq: centralized_eq}, $({F^T}R^{-1}F)^{-1}$ should be invertible. In order to calculate \eqref{eq: centralized_eq} we need to access all data in the system by a single (or centralized) control unit. As mentioned before, issues like communication bottleneck, data privacy and cybersecurity, are the main reasons which leads power system to utilize decentralized approaches.

\subsubsection{Matrix splitting}
In order to obtain SE problem's solution in a distributed manner, one can use matrix splitting method and after doing a certain number of iterations the answer converges to the centralized solution \cite{minot2015fully}. The main equation of matrix splitting for a problem of $Ax=y$ is:
\begin{equation}
\label{eq: matrixSplit}
x^{t+1} = M^{-1}Nx^{t}+M^{-1}y
\vspace{-0.3 cm}
\end{equation}
that $A$ is written as the sum of an invertible (or diagonal) matrix $M$, and a matrix $N$; i.e. $A = D+E$, or $A=M+N$ so that $M=D+E'_{ii}$ and $N=E-E'_{ii}$. Note that, $D$ contains diagonal arrays and $E$ contains off-diagonal arrays of matrix $A$. And $E'_{ii}$ is a diagonal matrix which is defined as follows:
\begin{equation}
\vspace{-0.3cm}
E'_{ii} = \alpha\times\sum_{j=1}^{n}|{E_{ij}}|
\vspace{-0.3 cm}
\end{equation}
that we have assumed $\alpha = 1$ for simplicity. It is to be noted that, \eqref{eq: matrixSplit} converges if the spectral radius of $M^{-1}N$ matrix be less than 1 ($\rho(M^{-1}N) < 1$). Using \eqref{eq: matrixSplit} iteratively, leads to convergence to the system $Ax=y$ final solution, i.e. $x^*$.
\subsubsection{Gossip based}
Here another approach to solve \eqref{eq: centralized_eq} in a distributed manner is presented, which is discussed in \cite{frasca2015distributed} that the authors have considered measurement units in an asynchronous manner (i.e. gossip communication protocol).

Considering the DC approximation, the SE problem in the least squares setting can be formulated by \eqref{eq: main_objective_function} and \eqref{eq: centralized_eq}. Based on what has been stated in \eqref{eq: centralized_eq}, this problem has a closed-form solution. Let's assume, $L = F^TR^{-1}F$ and $u = F^TR^{-1}z$. One way to compute this solution $x^*$ is through the gradient based iterative algorithm given by:
\begin{equation}
\label{eq: gossip based}
    x(k+1)=(I-\tau L)x(k)+\tau u
    \vspace{-0.2cm}
\end{equation}
And the parameter $\tau$ is selected from the interval (0 , $2||L||^{-1}$); such a $\tau$ guarantees the matrix $I-\tau L$ to be Schur stable (i.e. the iterative method converges). At each iteration, a set of pair areas, randomly (based on uniform probability distribution) will be selected to update the common variables.
\subsubsection{Decomposition method}
In this part the method provided in \cite{conejo2007optimization} is discussed. This method applies explicitly power flow and power injection equations to solve multi-area DC SE problem:
\begin{IEEEeqnarray}{c}
\label{eq: decomposition}
     min \text{ } F_k(x_k)+\sum_{l\in\Omega_k}{F_{kl}(x_k,\Tilde{x}_{l})}\nonumber\\
     F_k(x_k)= \displaystyle\sum_{i\in\Omega_k^P}{\omega_{k,i}^P(P_{k,i}^{m}-P_{k,i})^2}\nonumber\\
    + \displaystyle\sum_{(i,j)\in\Omega_k^{PF}}{\omega_{k,ij}^{PF}(P_{k,ij}^{m}-P_{k,ij})^2}\nonumber\\
     F_{kl}(x_k,\Tilde{x}_{l})= \displaystyle\sum_{i\in\Omega_{kl}^P}{\omega_{kl,i}^P(P_{kl,i}^{m}-P_{kl,i})^2} \\
    + \displaystyle\sum_{(i,j)\in\Omega_{kl}^{PF}}{\omega_{kl,ij}^{PF}(P_{kl,ij}^{m}-P_{kl,ij})^2}\nonumber\\
    +\displaystyle\sum_{i\in\Omega_{kl}}{\omega_{l,i}^{x}(\Tilde{x}_{l,i}-x_{l,i})^2}, \nonumber
\end{IEEEeqnarray}
where $F_{kl}$ is weighted measurement error function for area \textit{k} involving state variables of area \textit{k} and \textit{l}, $F_k$ is weighted measurement error function for area \textit{k} involving only state variables of area \textit{k}, and $\Omega_k$ is the set containing indices for all neighboring areas of area \textit{k}, $\omega$ is weighting factor, $P_{(.),i}^{(m)}$ is active power injection measurement at bus $i$, $P_{(.),ij}^{(m)}$ is active power flow measurement in between bus $i$ and $j$; $P_{(.),i}$ and $P_{(.),ij}$ are the physical equation related to power injection and power flow, respectively. 

In order to solve \eqref{eq: decomposition}, MATLAB solver (Sequential quadratic programming (SQP)) via MATLAB \textit{R2018b} has been applied.

\subsubsection{ADMM}
In \cite{kekatos2013distributed} a new method has been developed  for solving DSE, which is based on ADMM \cite{boyd2004convex}. As claimed by the authors, ADMM increases existing SE solvers performance and convergence of the method to its centralized counterpart is guaranteed even if we don't have local observability. ADMM can also be considered in the same category as decomposition methods, but due to multiple applications of this method recently, we have decided to consider it separately.

In general, the DSE problem can be formulated as:
\begin{IEEEeqnarray}{C}
    \min_{x_k} \displaystyle\sum_{k=1}^{K}{f_{k}(x_k)} \nonumber \\
    x_k[l]=x_l[k]$ , $\forall l \in N_k$ , $\forall k
\end{IEEEeqnarray}
where $N_k$ is the set of areas sharing states with area \textit{k} and $x_{k,l}$ is auxiliary variable introduced per pair of interacting areas \textit{k}, \textit{l}.

The constraint forces neighboring areas to consent on their shared variables. Augmented Lagrangian function is as follows:
\begin{IEEEeqnarray}{c}
    L(\{x_k\},\{x_{kl}\};\{v_{kl}\}) \nonumber \\
    :=\displaystyle\sum_{k=1}^{K}[f_{k}(x_k)+\sum_{l \in N_k}(v_{k,l}^T(x_{k[l]}-x_{kl})+ \\
    \frac{c}{2}||x_{k[l]}-x_{kl}||_2^2)],\nonumber
\end{IEEEeqnarray}

where $v_{k,l}$ is Lagrangian multiplier and $c>0$.
\begin{IEEEeqnarray}{C}
    \{x_k^{t+1}\}:= \textit{ arg min } L(\{x_k\},\{x^t_{kl}\};\{v^t_{kl}\}) \nonumber \\
    \{x_{kl}^{t+1}\}:= \textit{ arg min } L(\{x_k^{t+1}\},\{x_{kl}\};\{v^t_{kl}\}) \\
    v_{k,l}^{t+1}:=v_{k,l}^t+ c(x^{t+1}_{k[l]}-x^{t+1}_{kl}), \text{  } \forall k , \nonumber
\end{IEEEeqnarray}

\subsection{Convergence criterion}
One of the trivial ways to stop an algorithm is to set specific number of iterations and hand out the solution when the iterations finish. Obviously, this way can not give a satisfactory result to most problems, specially SE which plays a vital role in power system management. In addition, the problem is not centralized anymore, which hands out the fact that we need to develop and implement a simple yet effective distributed method to deal with it. The following algorithm shows the general approach to optimal partitioned DSE with proposed convergence criterion.

\begin{algorithm}[H]
\begin{algorithmic}

\label{al: convcriterion}
\caption{DSE with convergence criterion}

\State - Optimal partitioning of the system
\State - Initialization of the DSE parameters 
\State - Define area number (AN), state number (SN), each area's measurements and needed data
\State - Specify convergence criterion parameter ($\epsilon$)
\State - Do the first iteration and then transmit the needed data between each area
\While { $||x^{t} - x_{i,k}^{t-1}|| > \epsilon $ }
\State Doing local computation
\For {$k = 1$ to AN}
\For {$i = 1$ to SN}
\If {$||x_{i,k}^{t} - x_{i,k}^{t-1}|| < \epsilon $}
\State $x_{i,k}$'s in the next steps will be equal to $x_{i,k}^{t-1}$
\State No need to transfer this data anymore
\Else 
\State keep on sending the needed data
\EndIf
\EndFor
\EndFor
\EndWhile
\end{algorithmic}
\end{algorithm}

\subsection{Power system partitioning}
It is possible to represent the entire power system using an undirected weighted graph and the connectivity between vertices (buses) of this graph (the power system) can be represented by the following connection matrix ($C_L$):
\begin{equation}
\label{eq: connection-mat}
C_L = \begin{bmatrix}
 c_{1,1} & \hdots & c_{1,M}\\ 
 \vdots & \ddots & \vdots\\ 
c_{M,1} & \hdots & c_{M,M}
\end{bmatrix}
\end{equation}

\begin{IEEEeqnarray}{l}
    s.t. \nonumber \\
    c_{i,j} = c_{j,i}, \hspace{0.5 cm}\{i,j\} = 1, 2, ..., M \nonumber \\
    c_{i,i} = 0 \nonumber
\end{IEEEeqnarray}
where $M$ is number of buses and the availability of a physical connection between nodes $i$ and $j$. So, if there is a connection between nodes $i$ and $j$, the value of $c_{i,j}$ will be assigned $"1"$, else it would be $"0"$.

We need to define a weight matrix ($W_L$) with value ($w_{i,j}$) for each element corresponding to connection matrix that introduced in \eqref{eq: connection-mat} such as:

\begin{equation}
\label{eq: connection-mat-weight}
    W_L = \begin{cases}
    w_{i,j},& \text{if } c_{i,j} = 1 \text{ and } i\neq j\\
    0,              & \text{if } i = j
\end{cases}
\end{equation}

\begin{equation*}
\{i,j\} = 1, 2, ..., M
\end{equation*}

Based on what has been mentioned in \eqref{eq: connection-mat} and \eqref{eq: connection-mat-weight} the total cost ($TC_L$) for cutting the connection between buses $i$ and $j$, can be obtained using $TC_L(i,j) = c_{i,j}w_{i,j}$. Finally, if we want to divide a power system with $M$ bus to $K$ areas, we can formulate the objective function ($J_k$) for each area (or partition) of the system as follows:
\begin{IEEEeqnarray}{c}
\label{eq: partition-obj-fun}
   min \hspace{1 mm} J_k = min \hspace{1 mm} \sum^{M}_{i = 1}\sum^{M}_{\substack{j = 1 \\ j\notin\phi_k}}{TC_L(i,j)}
\end{IEEEeqnarray}
\begin{IEEEeqnarray}{l}
    s.t. \nonumber \\
n(\phi_k) > b_{lim} \nonumber \\ 
K \geq 2 \nonumber
\end{IEEEeqnarray}
where $k$ indicates number of area ($k = 1, 2, …, K$); $i$ and $j$ indicate the bus number ${i , j} = 1, 2, …, M$; $\phi_k$ is the set of buses in area $k$; $n(\phi_k)$ and $b_{lim}$ are the number of elements in $\phi_k$ and minimum number of bus we expect to be in each area, respectively; 
It is to be noted that the specified constraints in \eqref{eq: partition-obj-fun} make sure that the number of buses in each area are more than a pre-specified threshold. Additionally, considering $K \geq 2$, avoids having only one area which is same as CSE.

\section{Simulation results and discussion} \label{sec: results}

\begin{figure}[h!]
\centerline{\includegraphics[width={60 mm}]{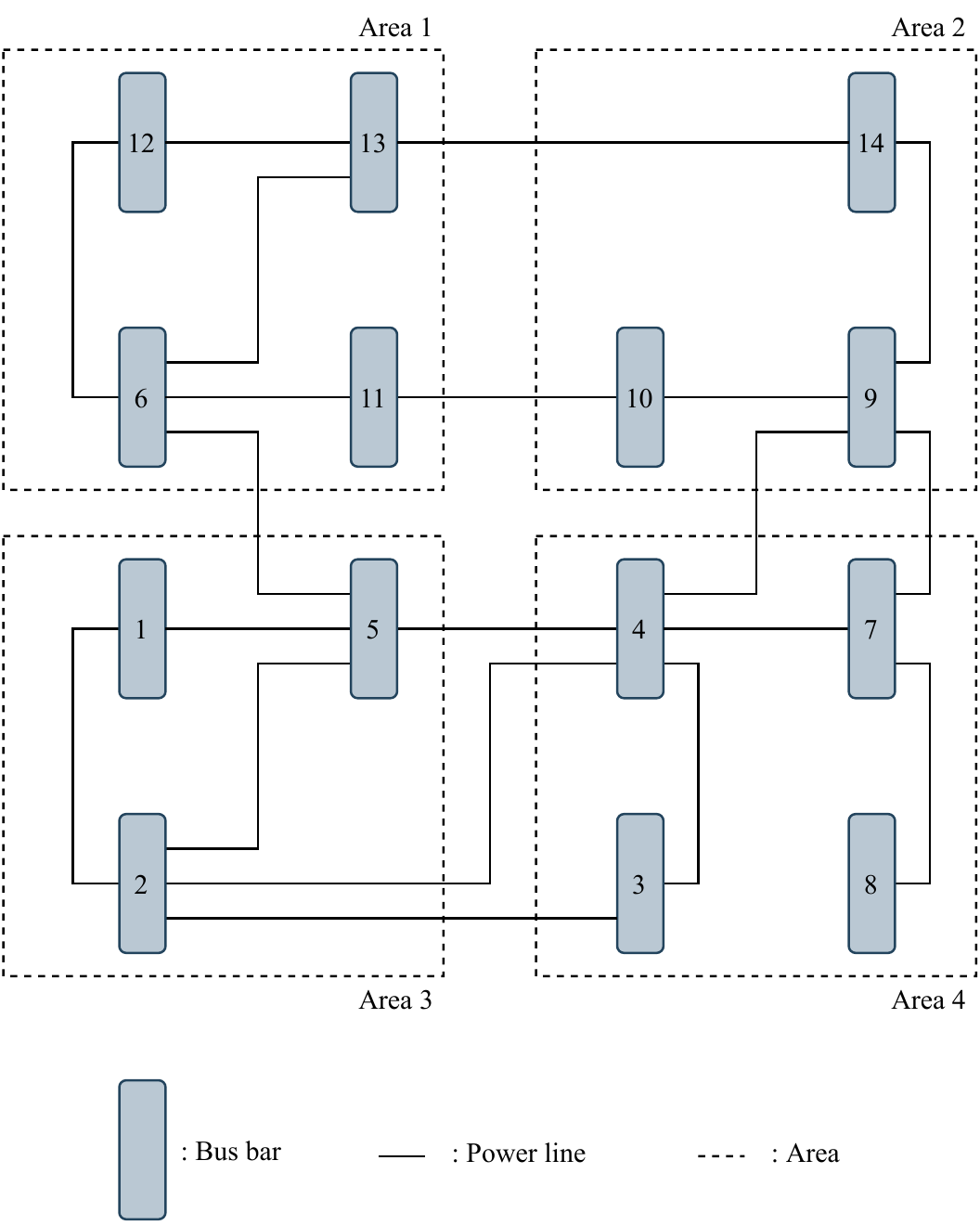}}
\caption{Topology of the IEEE 14 bus system}
\label{Fig:IEEE14SysTop}
\end{figure}

In this section, the proposed method's results on a test case, i.e. IEEE 14 bus system, are presented. The system has been divided into four areas. Fig. \ref{Fig:IEEE14SysTop} shows the topology of the studied test case.

As mentioned in section \ref{sec: prob-form}, static DC DSE has been considered in this study, where the state variables would be only phase angles at each bus. It is to be noted that the measurements are consist of active power flows and injections. The noise covariance for all measurement units has been considered $10^{-4}$, and the initial value for state variables are “0”. Moreover, bus number “1” has been selected as the slack bus. 
The simulation has been implemented via MATLAB $R2018b$ on a computer with Intel(R) Core i5 processor and 8 GB of RAM.

Table \ref{tab: num-res-14} provides the detailed numerical results of DSE for IEEE 14 bus system. It is to be noted that system data and area specification for IEEE 14 bus system is adapted from \cite{kekatos2017psse}. The DC centralized state estimation objective value for IEEE 14 bus system is $10.0524$. 

Also, Table \ref{tab: num-res-118} presents numerical results for IEEE 118 bus system. For the sake of brevity, the distributed scheme of IEEE 118 bus system is not provided here but it should be mentioned that, the topology of distributed IEEE 118 bus system is adopted from \cite{xia2019distributed}. The objective value for IEEE 118 bus system is $102.7758$. The main reason for considering IEEE 118 bus system is to check scalability of the problem. Additionally, in contrast to the IEEE 14 bus system, in bulk power systems, like IEEE 118 bus, the effects of considering convegence criterion are more visible.

\begin{table}[h!]
\centering
\caption{Numerical results of IEEE 14 Bus system}
\label{tab: num-res-14}
\resizebox{\columnwidth}{!}{%
\begin{tabular}{||c|c|c|c|c|c|c|c||}
\hline
Methods                           &      & Iter & $\epsilon_1$      & $\epsilon_2$      & CB      & OT      & OV\\ \hline 
\multirow{2}{*}{Matrix splitting} & WOCC & 1042 & 1.26e-3  & 1.31e-4  & 8.4546  & 529.45  & 10.0565 \\  
                                  & WCC  & 927  & 4.6e-3   & 4.16e-4  & 8.6483  & 472.15  & 10.1307 \\ \hline
\multirow{2}{*}{Gossip based}     & WOCC & 2217 & 2.59e-3  & 3.15e-4  & 0.56    & 1109.06 & 10.0689 \\  
                                  & WCC  & 1870 & 1.27e-2  & 1.22e-3  & 0.87308 & 935.873 & 10.5662 \\ \hline
\multirow{2}{*}{Decomposition}    & WOCC & 40   & 4.14e-3  & 5.61e-4  & 2.83    & 22.83   & 10.5615 \\ 
                                  & WCC  & 40   & 4.14e-3  & 5.61e-4  & 2.87    & 22.87   & 10.5615 \\ \hline
\multirow{2}{*}{ADMM}             & WOCC & 245  & 2.39e-2  & 2.43e-3  & 0.42828 & 122.93  & 12.036  \\ 
                                  & WCC  & 213  & 2.3e-2   & 2.37e-3  & 0.38293 & 106.88  & 11.8174 \\ \hline
\end{tabular}%
}
\end{table}

\begin{table}[h!]
\centering
\caption{Numerical results of IEEE 118 Bus system}
\label{tab: num-res-118}
\resizebox{\columnwidth}{!}{
\begin{tabular}{||c|c|c|c|c|c|c|c||}
\hline
Methods                           &      & Iter      & $\epsilon_1$ & $\epsilon_2$      & CB  & OT        & OV\\ \hline
\multirow{2}{*}{Matrix splitting} & WOCC & 65301     & 2.95 & 6.3e-2   & 7319.3546     & 39969.85    &   396.283 \\
                                  & WCC  & 39744     & 7.19         & 7.6e-2  & 3910.7458   & 23782.75 & 2194.4171  \\ \hline
\multirow{2}{*}{Gossip based}     & WOCC & 58811     & 17.0111      & 0.30187   & 98.523      & 29504.02 & 12130.307 \\
                                  & WCC  & 37163     & 17.8056      & 0.33289   & 100.0473    & 18681.55  & 42593.4768 \\ \hline
\multirow{2}{*}{Decomposition}    & WOCC & 190       & 5.42e-2     & 2.71e-3 & 149.8563    & 244.86    & 109.6939   \\
                                  & WCC  & 190       & 5.42e-2    & 2.71e-3 & 149.8563    & 244.86    & 109.6939   \\ \hline
\multirow{2}{*}{ADMM}             & WOCC & 1621      & 1.11         & 1.6e-2  & 5.8758      & 816.37    & 142.5793   \\
                                  & WCC  & 998       & 1.01         & 1.4e-2  & 3.9752      & 502.98 & 137.014    \\ \hline
\end{tabular}
}
\end{table}

Results provided in Table \ref{tab: num-res-14} and \ref{tab: num-res-118} are separated into two different categories. First one is without modified convergence criterion (WOCC) and the second on is with modified convergence criterion (WCC). Also, the number of iterations (Iter) of different methods, error values compared to centralized solution, computational burden (CB) and overall elapsed time (OT) in seconds (sec), and finally the objective function value (OV) are described for both categories here. Convergence limit $\epsilon$ was set to $10^{-6}$ for all cases. Two different scales were applied for measuring the error of each method’s solution compared to the answer obtained using the centralized method. $\epsilon_1$ is the sum of absolute values of difference between centralized and distributed solution (i.e. $\sum|x_{cent} - x_{dist}|$), and $\epsilon_2$ is $max(|x_{cent} - x_{dist}|)$. Computation burden means the time that has been spent by computer to solve the problem in a distributed manner. As stated in \cite{glavic2013tracking}, time delay for data transmission in power system can be considered between $0.1$ to $0.5$ (sec). So, data transmission delay $tdelay = 0.5$ (sec) as the worst case, and the overall time can be calculated using the following equation:
\begin{equation}
OT = (tdelay \times Iter ) + CB
\end{equation}
Finally, the OV for optimal state variables, which was obtained applying different methods, was evaluated using \eqref{eq: main_objective_function}.

It is to be noted that, simulation has not been done in parallel, but on a single computer. So, the time represented in \ref{tab: num-res-14} and \ref{tab: num-res-118} are sum of the time spent in all 4 or 6 areas for IEEE 14 and IEEE 118 bus system, respectively. In order to select the best algorithm amongst the ones which have been presented, features such as scalability, data needed to be transmitted and closer objective value to the centralized solution. Taking into account the mentioned details, the decomposition methods serves the best for the purpose of DSE. 

After specifying the DSE algorithm that has the closest results to CSE, the proposed optimal system partitioning has been applied on IEEE 14 bus system. For this purpose, $M = 14$; $b_{lim} = 3$; and $w_{i,j} = 0.01 \hspace{2 mm} \forall \hspace{1 mm} i,j$; And MATLAB solver (Sequential quadratic programming (SQP)) has been applied for solving \eqref{eq: partition-obj-fun}.

\begin{figure}[h]
    \centering
    \includegraphics[width=\linewidth]{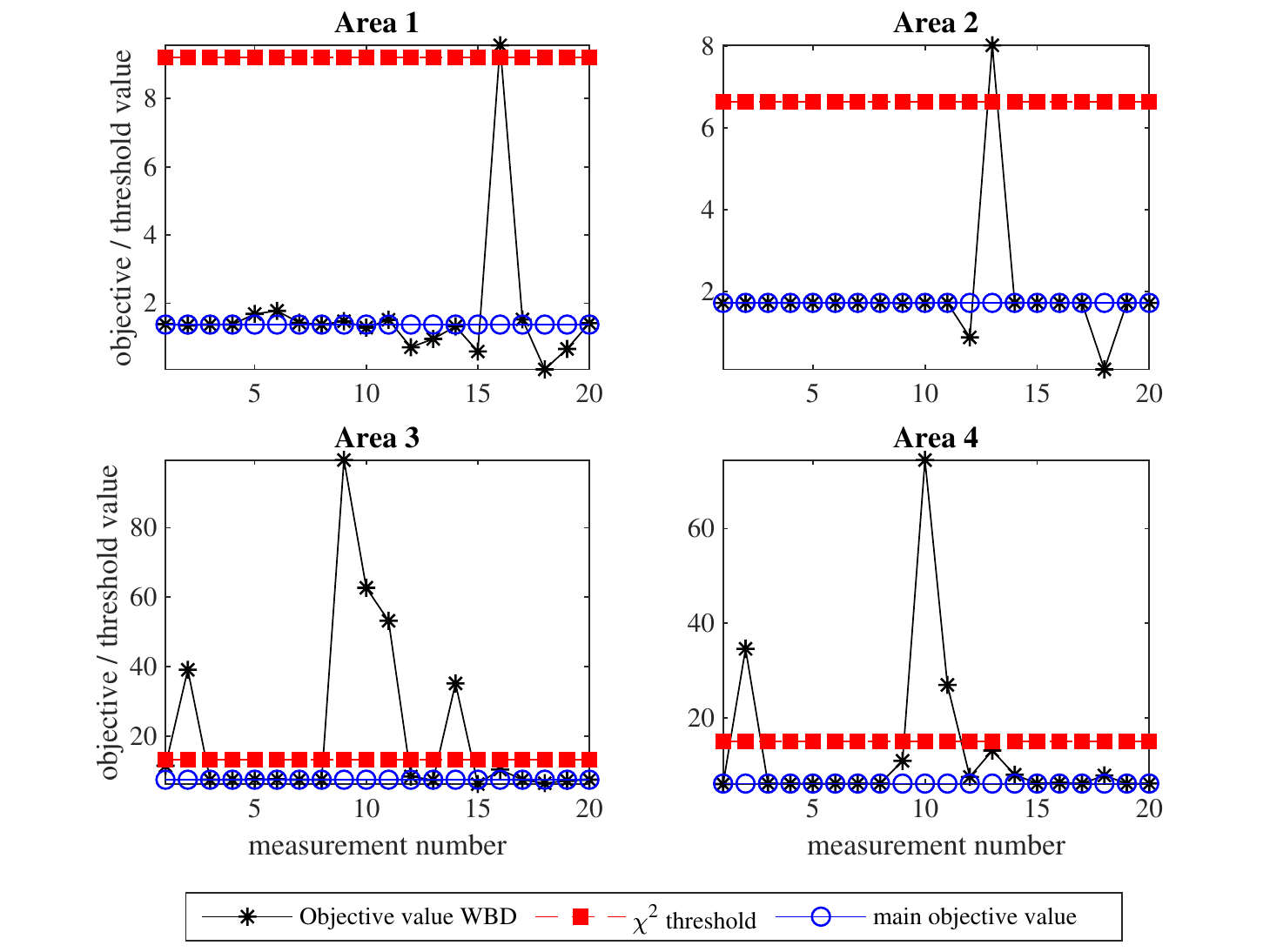}
    \caption{Objective value of areas by increasing each measurement value 10\% (each at a time) for partitioning case 1}
    \label{Fig: norm-part}
\end{figure}

\begin{figure}[h]
    \centering
    \includegraphics[width=\linewidth]{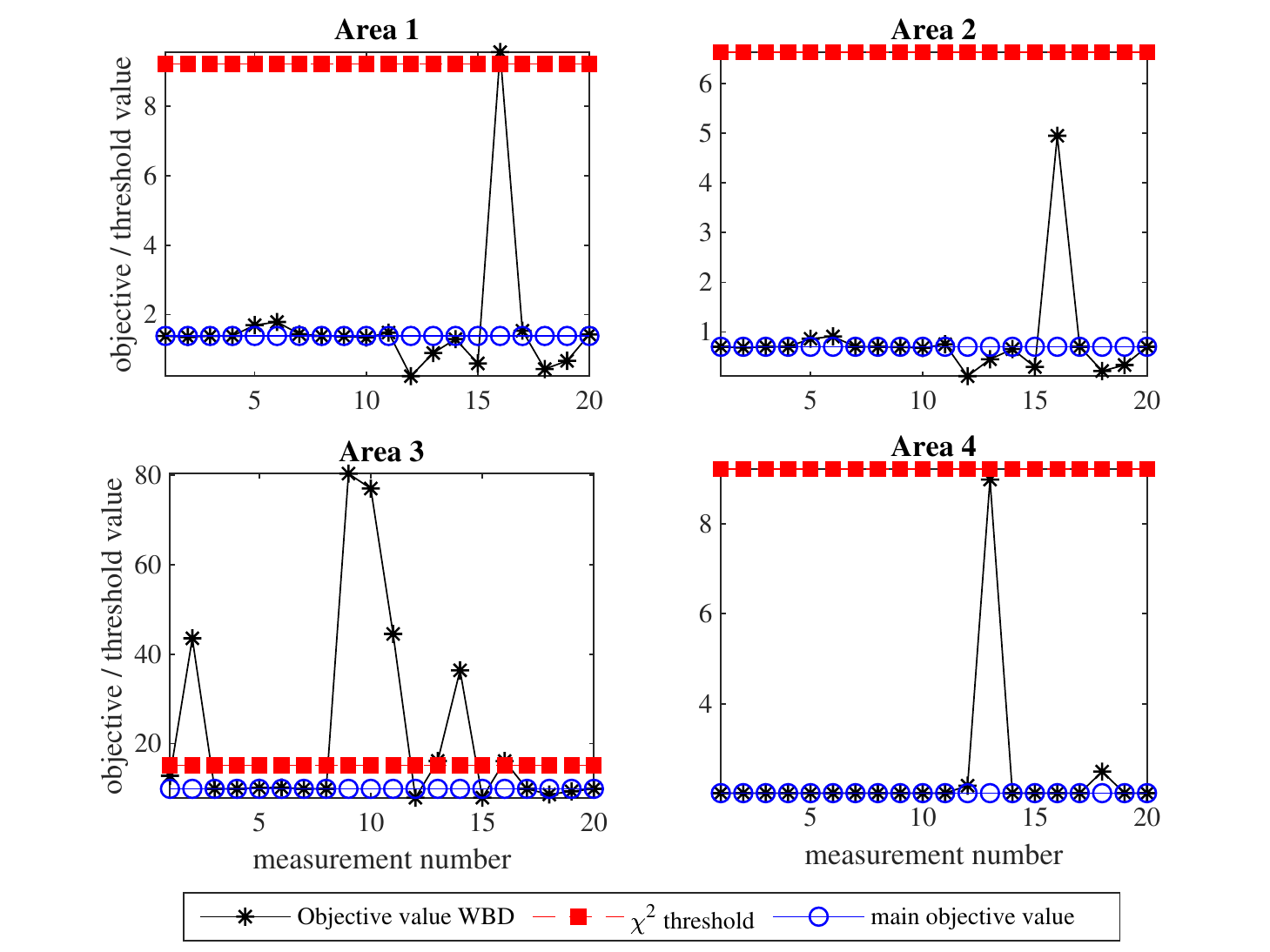}
    \caption{Objective value of areas by increasing each measurement value 10\% (each at a time) for partitioning case 2}
    \label{Fig: dif-part}
\end{figure}

In order to check the security of the system, in Fig. \ref{Fig: norm-part} and \ref{Fig: dif-part} a sensitivity analysis on the measurements has been done. The value of the measurement unit, has been increase by $10\%$ each separately and the results of all areas have been collected in the one figure. The aim is to identify objective value with bad data and compare it with the chi-square value (chi-square probability distribution function is conventionally used for bad data detection in power system).

The following figures show the results for two case. Fig. \ref{Fig: norm-part} is related to the partitioning which is normally used in the literature and the system configuration is as follows: Area 1 = \{6  11 12 13\}, Area 2 = \{14 9  10\}, Area 3 = \{1 2  5\}, Area 4 = \{3 4 7  8\}. The number of bad data detection is 7 in this case.

Fig. \ref{Fig: dif-part} is related to the proposed partitioning, the case when the system configuration is as follows: Area 1 = \{6  12 13\}, Area 2 = \{14 11  10\}, Area 3 = \{1 2  5 3 4\}, Area 4 = \{9 7 8\}. The number of bad data detection, similar to case 1, is 7 as well.

It is to be noted that, there might be a case, that two areas are going to have residuals more than the chi-square threshold (which means there is a bad data), at the same time. In this case it will be counted as one. Additionally, some of the measurements have zero value, so there is no change in their value, in whole 4 areas. It is clear that the overall security of the system has not changes in both cases has not changed.

\begin{table}[h!]
\centering
\caption{Numerical results for comparing case 1 and case 2}
\label{tab: comp-cas1-cas2}
\begin{tabular}{ccccc}
        &  Iteration & $\sum J_k$ & DSE Objective  & Error (\%) \\
 case 1 &  42 & 0.18 & 10.56  & 4.85 \%  \\
 case 2 &  37 & 0.14 & 10.23  & 1.76 \%
\end{tabular}%
\end{table}

Finally, Table \ref{tab: comp-cas1-cas2} compares the numerical results for case 1 and 2. Second case, which is related to system optimal partitioning has led to less iteration number that consequently results in less data communication and faster implementation. Then the sum partitioning objective value for all areas ($\sum J_k$) is presented. The individual partitioning objective has been evaluated for each area using \eqref{eq: partition-obj-fun}, to compare between partitioning scheme available in the literature and the proposed one. The obtained $\sum J_k$ result for case 2 it better than case 1. At the same time, due to decreased number of auxiliary variables due optimal partitioning, case 2 has lower objective value compared to case 1, which is closer to the centralized solution. And, finally, the error percentage that shows the relative error of DSE objective compared to CSE  objective value that is $10.0524$.
\section{Conclusion} \label{sec: conclusion}
In this paper we presented a modified convergence criterion for DSE application considering features such as iteration number, convergence rate and needed data to be transmitted between areas. After that an optimal partitioning method which maintains the security of the system while decreases the number of auxiliary variables of the DSE problem was introduced. 

Based on the obtained results, application of the modified convergence criterion will decrease the number of iterations to a high extent. Additionally, the proposed partitioning method is effective in case of decreasing the number of auxiliary variables of the DSE problem, and consequently helps to reach to an optimal point closer to CSE. 

Optimally selecting number of areas, measurements (specially considering lack of measurements) and buses in each area simultaneously, for the purpose of DSE can be considered as a potential future research direction.

\bibliographystyle{IEEEtran}
\bibliography{bibliography.bib}

\end{document}